# Evolution of dense star clusters

G. S. Bisnovatyi-Kogan

*Institute for Space Research, Academy of Sciences of the USSR, Moscow*

A qualitative study is made of the evolutionary tracks of various dense star clusters, with allowance for the evaporation of stars and for head-on collisions, until relativistic collapse occurs or the cluster breaks up.

In a system of point masses, two-body collisions will tend to establish a Maxwellian distribution, and high-velocity objects will escape from the system. Following Ambartsumyan[1] and Spitzer,[2] let us take

$$v_{esc} = 4\bar{v}^2 = 6kT,$$

where $\bar{v}$ is the mean velocity of the stars, $v_{esc}$ is the escape velocity, and $T$ is the temperature parameter specifying the Maxwellian distribution. We then obtain for the variation in the number $N$ of members of the system the equation[1,2]

$$\frac{dN}{dt} = -0.0074 \frac{N}{\tau} = -8.4 \cdot 10^{-9} \left(\frac{mN}{\bar{R}^3}\right)^{1/2} (\ln N - 0.45). \quad (1)$$

Here $m$ denotes the mass of a star in solar masses and $\bar{R}$ is the mean radius of the cluster in parsecs. The relaxation time $\tau$ with respect to Coulomb collisions is given[3] by

$$\tau = 8.8 \cdot 10^5 \sqrt{\frac{N\bar{R}^3}{m}} \frac{1}{\ln N - 0.45} \text{ sec}. \quad (2)$$

For interactions taking place by an inverse square law, distant encounters with a small momentum transfer will predominate. The dispersing particles carry nearly

zero energy, so that the energy E of the system remains approximately constant[4]:

$$E = -\frac{1}{4}\frac{GM^2}{\overline{R}} = -\frac{1}{4}\frac{G(mN)^2}{\overline{R}} \doteq \text{const.} \quad (3)$$

In an isolated system, low-energy particles traveling out into low-density regions near the periphery will collide very rarely; hence close encounters between stars in the dense central regions of the cluster, leading to an escape of particles with finite energy, may play an important role.[5-7] These collisions will occur more rarely by a factor of $\approx \ln N$ if the mean density is the same. On the other hand, tidal interactions between globular clusters and the stars in a galaxy, or between one galaxy and another, will cause particles to escape with negative energy. We shall adopt Eq. (3) for the rough evolutionary estimates given below.

As Agekyan[8] has shown, a cluster with little initial flattening will tend to become globular, so that rotation may be neglected in the evolution of such clusters. Spitzer and Saslaw[9] have examined the evolution of a rotating cluster in detail. As stars evaporate from the cluster and the remnant flattens, head-on collisions will become increasingly important, causing stars to coalesce and gas to be ejected. The time $t_c$ intervening between such

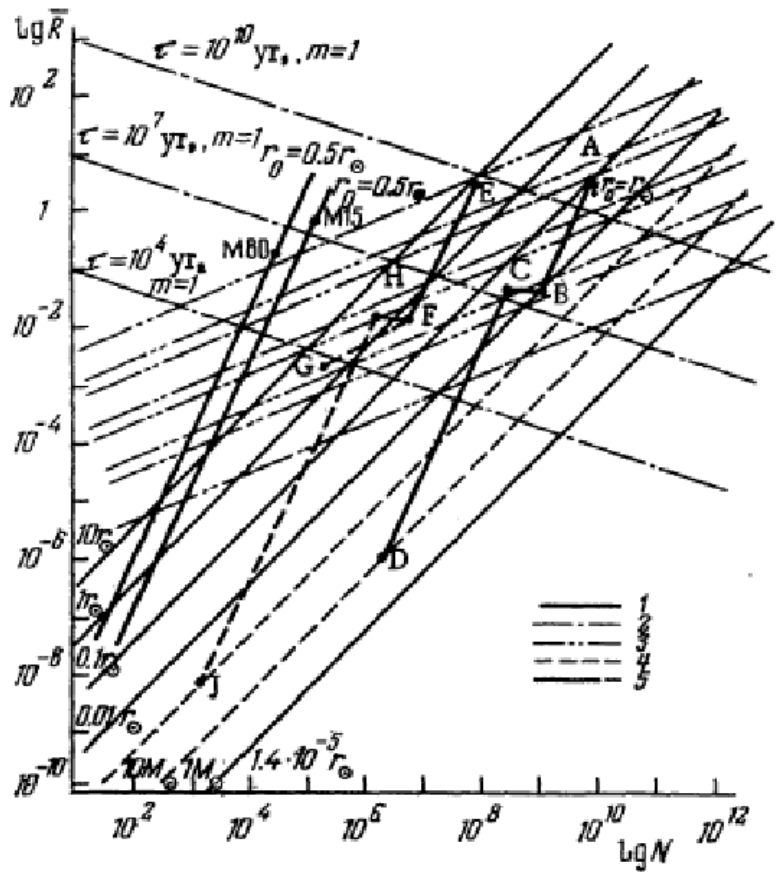

FIG. 1. Properties and evolutionary tracks of dense globular clusters, plotted in a ($\log \bar{R}$, $\log N$) diagram. 1) Lines $\tau = t_c$ for stars of selected radii; 2) lines of constant $\tau$ for $m = M/M_\odot = 1$; 3) lines of constant $t_c$ [the triple intersection points of these lines with the lines of constant $\tau$ and with the lines $\tau = t_c$ shown in the diagram determine the corresponding values of $t_c$ and $r$ (the dependence on $m$ is weak, so that lines $\tau =$ const and $t_c =$ const are shown in the diagram only for $m = 1$)]; 4) lines of relativistic collapse for $m = 1$ and $10$; 5) cluster evolutionary tracks.

collisions is given[9] by

$$t_c = \frac{10^{23} \overline{R}^{7/2}}{N^{3/2}(m^{1/2}r^2)(1 + 8.8 \cdot 10^7 \overline{R}/rN)} \text{ sec},\quad (4)$$

where r is the radius of the star in solar radii.

When a head-on collision occurs the luminosity of the system will rise abruptly because of the rapid radiation of energy by the gas. In addition, the growth in the mass of the stars will accelerate their evolution and lead to supernova outbursts. The stars will progressively turn into neutron and collapsing objects, and the system will continue to remain active because of accretion and radiation by the pulsars.[10-13]

In globular clusters and in the nuclear regions of elliptical galaxies, the angular momentum is small and the effect of rotation on the evolution may be neglected. If head-on collisions are unimportant, stars will continue to evaporate until relativistic collapse sets in.[14] A cluster with a truncated Maxwellian velocity distribution will lose stability[14,15] when $R = \nu R_g$, with $\nu = 4-5$. In view of Eq. (3), the mass $M_c$ of a collapsing remnant whose initial mass was $M_0$ will be

$$M_c = -\frac{8\nu E}{c^2} = 2\nu \frac{GM_0^2}{\overline{R}c^2} = \nu\left(\frac{v_{esc}}{c}\right)^2 M_0, \quad v_{esc} = \sqrt{\frac{2GM_0}{\overline{R}}}. \quad (5)$$

Figure 1 displays the relationships among the principal parameters of spherical clusters and shows their approximate evolutionary tracks for various initial conditions. The line ABCD could represent an active galaxy nucleus or a quasar. At the point A, the collision interval $t_c < \tau$; stars will collide, their mass will rise, and the evolution will proceed more rapidly. Since $t_c = 10^9$ yr at this point, stars of low mass (M ≲ 2 $M_\odot$) will be able to complete their evolution to the white dwarf stage. Since the rise in mass and the drop in the number of stars will accelerate the evaporation process, a small segment is marked at the top of the line AB; the cluster will traverse a segment of this approximate length in a time of order $10^9$ yr as its stars turn into white dwarfs and gas flows out of its nucleus. Until point B is reached, the main process will be the evaporation of white dwarfs, but at point B collisions between these objects will begin to predominate. Along the segment BC, collisions between white dwarfs and supernova explosions will make this phase in the life of the cluster especially active. Its duration of ≈ $10^7$ yr is consistent with the lifetime of quasars and active galaxy nuclei. At point C the cluster will consist mainly of neutron and collapsed stars, and as these evaporate the cluster will collapse with a mass of $10^6$-$10^7$ $M_\odot$ toward the point D.

The line beginning at the point E can be compared with the evolutionary track of an ordinary nucleus of an elliptical galaxy. Ordinary stars with masses of 0.5-1 $M_\odot$ will evaporate from the cluster along the line EF. At point F, head-on collisions begin to predominate. Since $t_c$ ≈ $10^6$ yr at point E, only the most massive stars will have been able to involve fully, ending their evolution in collapse and turning into neutron stars or black holes (point H). Further evaporation can result in collapse of a mass of $10^3$-$10^4$ $M_\odot$ at point J. If at point F coalescence should occur more rapidly than evolution, then the process will end with the formation of a single supermassive star having M ≈ $10^5$-$10^6$ $M_\odot$ (the line DG). The evolution of such a supermassive star can terminate either by collapse or by explosion, and even a comparatively small angular momentum could prove decisive[16-18] for the fate of the superstar.[1)]

For the nuclear region of a globular cluster, evaporation will predominate until a very small value of N is reached. Evolutionary tracks are shown at the left in Fig. 1 for the globular clusters M80 and M15, which have the most compact nuclei.[20] When the evolution ends, 10-100 stars will remain, and they could form a single massive star or a massive close binary. A massive black hole with $M \approx 10^3 M_\odot$, as has been proposed[21,22] to explain the presence of x-ray sources in the central regions of globular clusters, would hardly be expected to occur there.

Our qualitative discussion of the evolutionary process has neglected two factors that might prove important, although they are very difficult to evaluate quantitatively. The first factor is the formation of binaries in clusters with a small number of members, $N \lesssim 10^3$-$10^4$. If binaries were to develop, the cluster would expand because of the energy released as the radii of the binary systems decrease due to collisions with individual stars. The influence of this factor has been assessed in numerical experiments[23] involving a small number of members ($\lesssim 10^3$) and in model calculations for collisions of stars with binaries.[24] When binaries are present, the mass of the collapsing remnant becomes smaller, and the development of a black hole can be averted if N is less than some value $N_{lim}$. For want of anything better we may take an estimate $N_{lim} \approx 10^4$, although there is little support for such a value.

The other factor is the growth in the mass of a 5-10 $M_\odot$ black hole of stellar nature due to accretion. This process has been discussed by Ozernoi[25] and Frank,[26] but the result is far from clear because of the uncertainty in the properties of the gas and the accretion rate in a dense cluster. In principle, this process could lead to the formation of a massive black hole ($> 10^3 M_\odot$) from a star having a mass $M = 5$-$10 M_\odot$.

---

[1] The evolution of a rotating stellar system whereby a supermassive star forms at the center has been considered by Gurevich.[19]

## References

[1] V. A. Ambartsumyan, "On the dynamics of open star clusters" [in Russian] Trudy Astron. Obs. Leningrad. Gos. Univ. 7 (Uchen. Zap. Leningrad. Gos. Univ. No. 22; Ser. Mat. Nauk No. 4), 19-22 (1938).

[2] L. Spitzer, "The stability of isolated clusters," Mon. Not. R. Astron. Soc. 100, 396-413 (1940).

[3] S. Chandrasekhar, Principles of Stellar Dynamics, Univ. Chicago Press (1942).

[4] L. E. Gurevich and B. Yu. Levin, "The evolution of systems of gravitating bodies" [in Russian], Dokl. Akad. Nauk SSSR 70, 781-784 (1950).

[5] T. A. Agekyan, "The probability of a stellar encounter with a given change in absolute value," Astron. Zh. 36, 41-53 (1959) [Sov. Astron. 3, 46-58 (1959)].

[6] M. Henon, "L'évasion des étoiles hors des amas isolés," Ann. Astrophys. 23, 668-677 (1960).

[7] I. V. Petrovskaya, in: Astronomy in 1966, Itogi Nauki [Results of Science], VINITI [All-Union Inst. Sci. Tech. Info.], Moscow (1968).

[8] T. A. Agekyan, "The evolution of rotating systems of gravitating bodies," Astron. Zh. 35, 26-36 (1958) [Sov. Astron. 2, 22-31 (1959)].

[9] L. Spitzer and W. C. Saslaw, "On the evolution of galactic nuclei," Astrophys. J. 143, 400-419 (1966).

[10] S. A. Colgate, "Stellar coalescence and the multiple supernova interpretation of quasistellar sources," Astrophys. J. 150, 163-192 (1967).

[11] N. S. Kardashev, "Pulsars and nonthermal radio sources," Astron. Zh. 47, 465-478 (1970) [Sov. Astron. 14, 375-384 (1970)].

[12] G. S. Bisnovatyi-Kogan and R. A. Syunyaev, "Galaxy nuclei and quasars as infrared emission sources," Astron. Zh. 48, 881-893 (1971) [Sov. Astron. 15, 697-707 (1972)].

[13] J. Arons, R. M. Kulsrud, and J. P. Ostriker, "A multiple pulsar model for quasistellar objects and active galactic nuclei," Astrophys. J. 198, 687-707 (1975).

[14] Ya. B. Zel'dovich and M. A. Podurets, "The evolution of a system of gravitationally interacting point masses," Astron. Zh. 42, 963-973 (1965) [Sov. Astron. 9, 742-749 (1966)].

[15] J. R. Ipser, "Relativistic, spherically symmetric star clusters," Astrophys. J. 158, 17-43 (1969).

[16] G. S. Bisnovatyi-Kogan, "Explosions of large stars," Astron. Zh. 45, 74-80 (1968) [Sov. Astron. 12, 58-62 (1968)].

[17] K. J. Fricke, "Dynamical phases of supermassive stars," Astrophys. J. 183, 941-958 (1973).

[18] K. J. Fricke, "Dynamical phases of rotating supermassive stars," Astrophys. J. 189, 535-542 (1974).

[19] L. E. Gurevich, "Supermassive stars formed by inelastic evolution of stellar systems," Astron. Zh. 47, 32-42 (1970) [Sov. Astron. 14, 25-32 (1970)].

[20] C. J. Peterson and I. R. King, "Observed radii and structural parameters in globular clusters," Astron. J. 80, 427-436 (1975).

[21] J. N. Bahcall and J. P. Ostriker, "Massive black holes in globular clusters?," Nature 256, 23-24 (1975).

[22] J. Silk and J. Arons, "On the nature of the globular cluster x-ray sources," Astrophys. J. 200, L131-L135 (1975).

[23] S. J. Aarseth and M. Lecar, "Computer simulations of stellar systems," Ann. Rev. Astron. Astrophys. 13, 1-21 (1975).

[24] J. G. Hills, "Encounters between binary and single stars and their effect on the dynamical evolution of stellar systems," Astron. J. 80, 809-825 (1975).

[25] L. M. Ozernoi, "Failure of the supermassive black hole concept?," Observatory 96, 67-69 (1976).

[26] J. Frank, "Supermassive black holes dethroned?," Observatory 96, 199-200 (1976).